\documentclass[11pt]{article}
\usepackage[textwidth=15.2cm,textheight=22cm]{geometry}
\usepackage{amsmath,amssymb}
\usepackage{latexsym}
\usepackage{multicol}
\usepackage{graphicx}
\usepackage{bm}
\allowdisplaybreaks[1]

\newcommand{\be}{\begin{equation}}
\newcommand{\ee}{\end{equation}}
\newcommand{\ba}{\begin{eqnarray}}
\newcommand{\ea}{\end{eqnarray}}
\newcommand{\bdm}{\begin{displaymath}}
\newcommand{\nbar}[1]{\overline{#1}}
\newcommand{\edm}{\end{displaymath}}

\newcommand\fr[1]{\frac{1}{#1}}

\newcommand{\rom}[1]{\uppercase\expandafter{\romannumeral #1\relax}}
\renewcommand{\d}{\partial}

\newcommand{\E}{E_{7(7)}}
\def\parp{\partial^+}

\def\d{\delta}

\def\ba{\bar A}

\def\m#1{\mathcal#1}
\def\beq{\begin{equation}}
\def\eeq{\end{equation}}

\newcommand{\nn}{\nonumber}

\newcommand{\ndt}{\noindent}

\def\bea{\begin{eqnarray}}
\def\eea{\end{eqnarray}}
\def\beas{\begin{eqnarray*}}
\def\eeas{\end{eqnarray*}}
\def\sla{\raise.15ex\hbox{$/$}\kern-.57em}

\def\parm{{\partial}_{-}}
\def\parp{\partial^+}

\def\spa#1.#2{\left\langle#1\,#2\right\rangle}
\def\spb#1.#2{\left[#1\,#2\right]}

\begin{document}

\begin{titlepage}
\begin{flushright}    
{\small $\,$}
\end{flushright}
\vskip 1cm
\centerline{\Large{\bf{Exceptional versus superPoincar\'e algebra}}}
\vskip 0.5cm
\centerline{\Large{\bf{ as the}}}
\vskip 0.5cm
\centerline{\Large{\bf{defining symmetry of maximal supergravity}}}
\vskip 1.5cm
\centerline{Sudarshan Ananth$^*$, Lars Brink$^\dagger$ and Sucheta Majumdar$^*$}
\vskip .5cm
\centerline{$*$\it {Indian Institute of Science Education and Research}}
\centerline{\it {Pune 411008, India}}
\vskip .5cm
\centerline{$\dagger$\it {Chalmers University of Technology}}
\centerline{\it {Dept of Fundamental Physics,}} 
\centerline{\it {S-41296 G\"oteborg, Sweden}}
\vskip 1.5cm
\centerline{\bf {Abstract}}
\vskip .5cm
We describe how one may use either the superPoincar\'e algebra or the exceptional algebra to construct maximal supergravity theories in the light-cone formalism. The $d=4$ construction shows both symmetries albeit in a non-linearly realized manner. In $d=11$, we find that we have to choose which of these two symmetries to use, in constructing the theory. In order to understand the other ``unused" symmetry, one has to perform a highly non-trivial field redefinition. We argue that this shows that one cannot trust counterterm arguments that do not take the full symmetry of the theory into account. Finally we discuss possible consequences for Superstring theory and M-theory.
\vfill
\end{titlepage}

\section{Introduction}
\ndt In an extraordinary effort the entire Lagrangian of the $(\mathcal N=8,d=4$) supergravity theory was constructed by Cremmer and Julia~\cite{Cremmer:1979up}. They found that it had, in addition to the full $N=8$ supersymmetry, an $\E$ duality symmetry among the vector fields which appears as a $\sigma$-model symmetry of the scalar fields. The origin of their calculation was the construction of the $d=11$ supergravity theory, achieved with Scherk~\cite{Cremmer:1978km}. They dimensionally reduced the $d=11$ theory to four dimensions and used clever field redefinitions to make both symmetries manifest. The exceptional symmetry which is the maximally non-compact version of the algebra was fairly unexpected. This symmetry acts only on the scalar and vector fields and seems to lie outside the spacetime symmetries. A superspace formulation in terms of a superfield that contained the full supermultiplet with many auxiliary fields was subsequently found by Brink and Howe~\cite{Brink:1979nt}. Using it, Howe and Lindstr\"om~\cite{Howe:1980th} argued that there could be possible counterterms at the seven-loop level. This was taken as an indication that the theory would not make sense quantum mechanically at some loop order, probably at a level lower even than seven. When superstring theory appeared, supergravity came to be seen as the low-energy limit and as such is not believed to be consistent.
 \vskip 0.3cm
\ndt In more recent work, Bern, Dixon and collaborators~\cite{Bern:2011qn} have established a program to  compute higher loop orders in four-graviton scattering. They have been able to conclude that these are indeed finite at least up to four-loop order and there are clear signs that these amplitudes are ``more" finite than suggested by discussions based on the superfield formalism~\cite{Brink:1979nt}. We will return to this issue later in the paper.
\vskip 0.3cm
\ndt A very different description of supergravity was initiated by Bengtsson, Bengtsson and Brink~\cite{Bengtsson:1983pg} which involved the construction of $(\mathcal N=8,d=4$) supergravity iteratively in the coupling constant, as a representation of the $\mathcal N=8$ superPoincar\'e algebra acting on a superfield containing exclusively the physical degrees of freedom. They showed that this was identical to the light-cone gauge formulation of the theory. By choosing light-cone gauge and eliminating all unphysical degrees of freedom one is led to the same Hamiltonian, which in this case is also one of the generators of the algebra. This approach had only been used to derive the three-point coupling and as such was incomplete. Even so, one can learn a lot about the theory in this formalism since the Hamiltonian is remarkably simple to that order. One must bear in mind that the  $(\mathcal N=8,d=4$) supergravity lagrangian is probably the most complicated lagrangian ever constructed. In component form, expanding around a flat vacuum, there are of the order of 5000 four-point couplings. The four-point coupling in the light-cone gauge version was eventually constructed in~\cite{Ananth:2006fh} although the resulting expression is not simple. However, most of the juice in the theory is contained in the three-point coupling and we will only consider this here.
\vskip 0.3cm 
\ndt A natural question to ask is how the $\E$ symmetry is manifested in the light-cone gauge formulation. A duality transformation in this formulation is equivalent to a field redefinition. In the component formulation, and in the covariant superfield formulation, the $\E$ symmetry acts only on the scalar and vector fields. In the light-cone formulation a transformation on the superfield transforms all the fields in the supermultiplet. Hence an $\E$ transformation would also transform the graviton and spinor fields. Furthermore, the typical representations of the $\E$ symmetry, the $\bf 56$ and the $\bf 133$ are indeed intimately connected to only the vector representation and the scalar one. These issues were addressed by Brink, Kim and Ramond~\cite{Brink:2008qc}. When implementing the light-cone gauge one solves for the unphysical fields and this procedure mixes up all the fields in terms of new fields which constitute the superfield. Hence all the fields will be transformed under the $\E$ symmetry. Furthermore this symmetry is non-linearly realized in the sense that while $SU(8)$ is linearly realized, $\E/SU(8)$ is not. The supermultiplet is a representation of the $\E$ symmetry and somehow the exceptional group can distinguish bosons and fermions, a classical problem when dealing with exceptional groups that do not have spinor representations.
\vskip 0.3cm
\ndt The supermultiplet is a non-linearly realized representation of both the supersymmetry algebra and the $\E$ symmetry  and both these symmetries may be used to derive the interacting theory as was shown in~\cite{Brink:2008qc}. What happens if we ``oxidize" the theory back to $d=11$? In this paper, we will do this in two ways, one which explicitly shows the supersymmetry and another that exhibits the $\E$ symmetry. We will argue that the eleven-dimensional theory is invariant under both these defining symmetries, even though one has to perform a field redefinition to go from one approach to the other. 
\vskip 0.3cm
\ndt In section 2 we describe the {($\mathcal N=8,d=4$) supergravity in light-cone superspace in quite some detail so
that in section 3 we will be able to perform the oxidation to $d=11$. As mentioned, we will do this in two different ways which each preserve the relevant symmetry. We will then discuss how the approaches are related.  In section 4 we discuss higher exceptional symmetries such as the $E_{8(8)}$ symmetry that has been found in {($\mathcal N=16, d=3$) supergravity. We argue that this symmetry should also be hidden in higher dimensions. We will not have anything to say on the even higher exceptional symmetries at this stage. In section 5 we discuss the relevance of our results to the discussions about counterterms, possible finiteness and how these results might influence discussions on M-Theory and Superstring theory.
\vskip 0.5cm

\section{($\mathcal N=8,d=4$) Supergravity in light-cone superspace}

\vskip 0.5cm

\noindent In this section, we briefly review the light-cone superspace fomulation of ($\mathcal N=8,d=4$) supergravity. We work with the metric $(-,+,+,+)$, and introduce the following light-cone coordinates and derivatives
\bea
\begin{split}
{x^{\pm}}&=&\frac{1}{\sqrt 2}\,(\,{x^0}\,{\pm}\,{x^3}\,)\ ;\qquad {\partial^{\pm}}=\frac{1}{\sqrt 2}\,(\,-\,{\partial_0}\,{\pm}\,{\partial_3}\,)\ , \\
x &=&\frac{1}{\sqrt 2}\,(\,{x_1}\,+\,i\,{x_2}\,)\ ;\qquad {\bar\partial} =\frac{1}{\sqrt 2}\,(\,{\partial_1}\,-\,i\,{\partial_2}\,)\ , \\
{\bar x}& =&\frac{1}{\sqrt 2}\,(\,{x_1}\,-\,i\,{x_2}\,)\ ;\qquad {\partial} =\frac{1}{\sqrt 2}\,(\,{\partial_1}\,+\,i\,{\partial_2}\,)\ .
\end{split}
\eea

\vskip 0.5cm

\subsection{The superfield}

\noindent All the physical degrees of freedom in the $\mathcal N=8$ theory are contained in a single superfield~\cite{Brink:1982pd} written in terms of complex Grassmann variables $\theta^m$ ($m=1\,\ldots\,8$ are $SU(8)$ indices)
\bea\label{superfield}
\begin{split}
\phi\,(\,y\,)\,=&\,\frac{1}{{\parp}^2}\,h\,(y)\,+\,i\,\theta^m\,\frac{1}{{\parp}^2}\,{\bar \psi}_m\,(y)\,+\,\frac{i}{2}\,\theta^m\,\theta^n\,\frac{1}{\parp}\,{\bar A}_{mn}\,(y)\ , \\
\;&-\,\frac{1}{3!}\,\theta^m\,\theta^n\,\theta^p\,\frac{1}{\parp}\,{\bar \chi}_{mnp}\,(y)\,-\,\frac{1}{4!}\,\theta^m\,\theta^n\,\theta^p\,\theta^q\,{\bar C}_{mnpq}\,(y)\ , \\
\;&+\,\frac{i}{5!}\,\theta^m\,\theta^n\,\theta^p\,\theta^q\,\theta^r\,\epsilon_{mnpqrstu}\,\chi^{stu}\,(y)\ ,\\
\;&+\,\frac{i}{6!}\,\theta^m\,\theta^n\,\theta^p\,\theta^q\,\theta^r\,\theta^s\,\epsilon_{mnpqrstu}\,\parp\,A^{tu}\,(y)\ ,\\
\,&+\,\frac{1}{7!}\,\theta^m\,\theta^n\,\theta^p\,\theta^q\,\theta^r\,\theta^s\,\theta^t\,\epsilon_{mnpqrstu}\,\parp\,\psi^u\,(y)\ ,\\
\,&+\,\frac{4}{8!}\,\theta^m\,\theta^n\,\theta^p\,\theta^q\,\theta^r\,\theta^s\,\theta^t\,\theta^u\,\epsilon_{mnpqrstu}\,{\parp}^2\,{\bar h}\,(y)\ .
\end{split}
\eea

\noindent The fields in the superfield are as follows. $h$ and $\bar h$: the two-component graviton, ${\bar \psi}_m$: the $8$  spin-$\frac{3}{2}$ gravitinos, ${\bar A}_{mn}$: the $28$ gauge fields with ${\bar \chi}_{mnp}$ being the corresponding $56$ gauginos and ${\bar C}_{mnpq}$ representing the $70$ scalar fields. Complex conjugation of the fields is denoted with a bar. These fields are all local in the coordinates  
\bea
y~=~\,(\,x,\,{\bar x},\,{x^+},\,y^-_{}\equiv {x^-}-\,\frac{i}{\sqrt 2}\,{\theta_{}^m}\,{{\bar \theta}^{}_m}\,)\ .
\eea

\ndt We note that all the unphysical degrees of freedom have been integrated out. The superfield $\phi$ and its conjugate $\bar\phi$ satisfy the chiral constraints
\be
d^m\,\phi\,(\,y\,)\,=\,0\;\; ;\qquad {\bar d}_n\,{\bar \phi}\,(\,y\,)\,=\,0\ ,
\ee
\noindent where
\bea
d^{\,m}\,=\,-\,\frac{\partial}{\partial\,{\bar \theta}_m}\,-\,\frac{i}{\sqrt 2}\,\theta^m\,\parp\;\; ;\qquad{\bar d}_n\,=\,\frac{\partial}{\partial\,\theta^n}\,+\,\frac{i}{\sqrt 2}\,{\bar \theta}_n\,\parp\ ,
\eea

\noindent and are further related through the ``inside-out" constraint
\bea \label{inside-out}
\label{io}
\,{\phi}\,=\,\frac{1}{4}\,\frac{{(d\,)}^8}{{\parp}^4}\,{\bar \phi}\ ,
\eea
\noindent where ${(d\,)}^8\,=\,d^1\,d^2\,\ldots\,d^8$; this constraint being unique to maximally supersymmetric theories.

\vskip 0.5cm

\subsection{SuperPoincar\'e Algebra in $d=4$}
We review here, the results of \cite {Bengtsson:1983pg}, starting with the construction of the generators of the SuperPoincar\'e algebra at light-cone time $x^+=0$. The kinematical generators are, \newline
$\\ \bullet$ the three momenta,
\be
p^+_{}~=~-i\,\partial^+_{}\ ,\qquad p~=~-i\,\partial\ ,\qquad \bar p~=~-i\,\bar\partial\ ,
\ee
\newline
$\bullet$ the transverse space rotation,
\be
j~=~x\,\bar\partial-\bar x\,\partial+ S^{12}_{}\ ,
\ee
with $S^{12}_{}~=~ \,\frac{1}{ 2}\,(\,{\theta^\alpha}\,{{\bar \partial}_\alpha}\,-\,{{\bar \theta}_\alpha}\,{\partial^\alpha}\,)\,+\frac{i}{4\sqrt{2}\,\partial^+}\,(\,d^\alpha\,\bar d_\alpha-\bar d_\alpha\,d^\alpha\,)$ which satisfies
\be
[\,j\,,\,d^\alpha_{}\,]~=~[\,j\,,\,\bar d^{}_\beta\,]~=~0\ .
\ee
\newline
$\bullet$ and the ``plus-rotations",
\be
j^+_{}~=~i\, x\,\partial^+_{}\ ,\qquad \bar j^+_{}~=~i\,\bar x\,\partial^+_{}\ .
\ee
\be
 j^{+-}_{}~=~i\,x^-_{}\,\partial^+_{}-\frac{i}{2}\,(\,\theta^\alpha_{}\bar\partial^{}_\alpha+\bar\theta^{}_\alpha\,\partial^\alpha_{}\,)\ ,
\ee
which obey
\bea
\begin{split}
[\,j^{+-}_{}\,,\,y^-_{}\,\,]~&=~-i\,y^-_{}\ , \\
{[\,j^{+-}_{}\,,\,d^\alpha_{}\,\,]}~&=~\frac{i}{2}\,d^\alpha_{}\ ,\qquad {[\,j^{+-}_{}\,,\,\bar d_\beta^{}\,]}~&=~\frac{i}{2}\,\bar d^{}_\beta\ ,
\end{split}
\eea
\noindent The dynamical generators are, \newline
$\\ \bullet$ the light-cone Hamiltonian,
\be
p^-_{}~=~-i\frac{\partial\bar\partial}{\partial^+_{}}
\ee
\newline
$\bullet$ and the dynamical boosts,
\bea
j^-_{}&=&i\,x\,\frac{\partial\bar\partial}{\partial^+_{}} ~-~i\,x^-_{}\,\partial~+~i\,\Big( \theta^\alpha_{}\bar\partial^{}_\alpha\,+\frac{i}{4\sqrt{2}\,\partial^+}\,(\,d^\alpha\,\bar d_\alpha-\bar d_\alpha\,d^\alpha\,)\Big)\frac{\partial}{\partial^+_{}}\,\ ,\cr 
\bar j^-_{}&=&i\,\bar x\,\frac{\partial\bar\partial}{\partial^+_{}}~ -~i\,x^-_{}\,\bar\partial~+~ i\,\Big(\bar\theta_\beta^{}\partial_{}^\beta+\frac{i}{4\sqrt{2}\,\partial^+}\,(\,d^\beta\,\bar d_\beta-\bar d_\beta\,d^\beta\,)\,\Big)\frac{\bar\partial}{\partial^+_{}}\,\ .
\eea
which satisfy
\be
[\,j_{}^-\,,\,\bar j^+_{}\,]~=~-i\,j^{+-}_{}-j\ ,\qquad [\,j^-_{}\,,\,j^{+-}_{}\,]~=~i\,j^{-}_{}\ .
\ee

\vskip 0.5cm

\noindent The supersymmetry generators are of two varieties~\cite{Bengtsson:1983pg}, the kinematical
\bea
q^m_{\,+}\,=\,-\,\frac{\partial}{\partial\,{\bar \theta}_m}\,+\,\frac{i}{\sqrt 2}\,\theta^m\,\parp ;\qquad {\bar q}_{\,+\,n}=\;\;\;\frac{\partial}{\partial\,\theta^n}\,-\,\frac{i}{\sqrt 2}\,{\bar \theta}_n\,\parp\ ,
\eea
\noindent and the dynamical ones
\bea
\label{dynsus}
\begin{split}
q_-^{\,m}\,\equiv\,&i\,[\,\bar j^-\,,\,q^m_{\,+}\,]\,=\,\frac{\bar \partial}{\parp}\,q^m_{\,+}\,  , \\
{\bar q}_{-\,n}\,\equiv\,&i\,[\,j^-\,,\,{\bar q}_{\,+\,n}\,]\,=\,\frac{\partial}{\parp}\,{\bar q}_{\,+\,n}\, .
\end{split}
\eea

\vskip 0.5cm

\subsection{The action to order $\kappa$}
\vskip 0.5cm

\noindent The ${\mathcal N}=8$ supergravity action to order $\kappa$ in terms of the superfield described above reads

\be
\label{n=8}
\beta\,\int\;d^4x\,\int d^8\theta\,d^8 \bar \theta\,{\cal L}\ ,
\ee
where $\beta\,=\,-\,\frac{1}{64}$ and
\bea
\label{one}
{\cal L}&=&-\bar\phi\,\frac{\Box}{\partial^{+4}}\,\phi\,-\,2\,\kappa\,(\,\frac{1}{{\parp}^2}\;{\nbar \phi}\;\;{\bar \partial}\,{\phi}\;{\bar \partial}\,{\phi}+\,\frac{1}{{\parp}^2}\;\phi\,\partial\,{\nbar \phi}\,\partial\,{\nbar \phi})\ .
\eea

\noindent The d'Alembertian is
\bea
\Box\,=\,2\,(\,\partial\,{\bar \partial}\,-\,\partial_+\,\parm\,)\ ,
\eea
\noindent $\kappa\,=\,{\sqrt {8\,\pi\,G}}$ and Grassmann integration is normalized such that $\int d^8\theta\,{(\theta)}^8=1$.
\vskip 0.5cm
\ndt In principle this is to be augmented with the full non-linear superPoincar\'e algebra to this order. The important generator is the dynamical supersymmetry generator 

\be\label{Q}
\bar Q_m{}^{(\kappa)} \phi=  \frac{1}{\parp}(\bar \partial \bar q_m \phi {\parp}^2 \phi - \parp \bar q_m \phi \parp \bar \partial \phi).
\ee
Note that we have suppressed the $+$ index on $q_+$ to make things easier to read. By taking the complex conjugate of this formula we obtain $Q^m{}^{(\kappa)} \bar \phi$. Using the ``inside-out" constraint  (\ref{inside-out}) we can then derive  $Q^m{}^{(\kappa)} \phi$ and $\bar Q_m{}^{(\kappa)} \bar \phi$. Commuting these will yield $P^-{}^{(\kappa)} \phi$ and its complex conjugate. These can also be obtained from the action by taking a functional derivative remembering that $\phi$ is a constrained field. We also have to construct the non-linear parts of the generators $J^-$ and $\bar J^-$. These can be found in~\cite{Bengtsson:1983pg}.

\subsection{The $\E$ symmetry of the theory}
\vskip 0.1cm
\ndt As mentioned in the introduction the $\E$ symmetry of the theory was discussed in~\cite{Brink:2008qc}.  We have to write the symmetry algebra in terms of $\E/SU (8)\times SU(8)$, where the $SU(8)$ is the linearly realized $R$-symmetry of the superfield. The key point in the derivation is the assumption that $\E/SU(8)$ transformations commute with the supersymmetry. This might sound strange since the $SU(8)$ transformations transform the supersymmetry generators and two $\E/SU (8)$ transformations commute to an $SU(8)$ transformation. It is possible though since the $\E/SU (8)$ transformations act non-linearly. The final form for the transformations to order $\kappa$ is
\begin{align}
\delta \phi=&~-\frac{2}{\kappa}\,\theta^{klmn}_{}\,\overline\Xi^{}_{klmn}\nn \\
&+\frac{\kappa}{4!}\,\Xi^{mnpq}_{}\frac{1}{\partial^{+2}}\left(\overline d_{mnpq} \frac{1}{\partial^+}\phi\,\partial^{+3}_{}\phi \, -\,4\,\overline d_{mnp} \phi\,\overline d_q\partial^{+2}_{}\phi \,+\, 3\,\overline d_{mn} {\partial^+}\phi\,\overline d_{pq}\partial^{+}_{}\phi \right),
\end{align}
where $\overline d_{m_1...m_n} = \bar d_{m_1}....\bar d_{m_n}$ and $\overline\Xi^{}_{klmn} = \frac{1}{2} \epsilon_{klmnpqrs} \,\Xi^{pqrs}$, a constant. The $\E/SU(8)$ transformation can be written in a more compact way by introducing a coherent state-like representation 

\begin{equation}\label{E}
\delta \phi~=~
-\frac{2}{\kappa}\,\theta^{ijkl}_{}\,\overline\Xi^{}_{ijkl}\,+\,
\frac{\kappa}{4!}\,\Xi^{ijkl}  \left(\frac{\d}{\d\eta}\right)_{ijkl}\frac{1}{\partial^{+2}}\left(e^{\eta \hat{\bar d}} \partial^{+3} \phi\, e^{-\eta \hat{\bar d}}\partial^{+3} \phi \right)\Bigg|_{\eta=0} + \m O(\kappa^2),
\end{equation}
where 

$$
\eta\hat{\bar d} = \eta^m\frac{\bar d_m}{\partial^+},~~{\rm and}~~\left(\frac{\d}{\d\eta}\right)_{ijkl} \equiv~ \frac{\d}{\d\eta^i}\frac{\d}{\d\eta^j}\frac{\d}{\d\eta^k}\frac{\d}{\d\eta^l}\ .
$$
It is quite remarkable that the supermultiplet is a representation of  $\E$. In~\cite{Brink:2008qc} it is shown how this may be used to construct the interaction terms of the theory. In this respect, the $\E$ symmetry is as fundamental a symmetry as supersymmetry. In the latter case, the commutation of two supersymmetries yields the Hamiltonian. In the  $\E$ case the Hamiltonian commutes with the  $\E$ symmetry. In both cases the transformations in our formalism are non-linear. By going back to a covariant formulation the supersymmetry becomes linear while we cannot have the  $\E$ transform linearly on the whole multiplet. However in the light-cone formulation we treat both symmetries on an equal basis.

\section{Oxidation of the ($\mathcal N=8,d=4$) Supergravity to $d=11$}
\vskip 0.1cm

\ndt In~\cite{Ananth:2005vg}}, it was shown that the interacting $({\mathcal N}=8,d=4)$ theory could be restored to its eleven-dimensional avatar without altering the superfield. The additional $SO(7)$ is introduced through real coordinates, ${x^m}$ and their derivatives $\partial^m$ ($m$ runs from $4$ through $10$). The chiral superfield remains unaltered, except for the added dependence on these extra coordinates. The super-coordinate $\theta$ is now regarded as an  $SO(7)$ spinor $\theta^\alpha$. The superPoincar\'e algebra clearly needs to be suitably enlarged. Let us first describe the free theory and then write the interactions in two different ways, first showing the explicit supersymmetry in the theory and then repeating the calculations from~\cite{Ananth:2005vg}} to display the $\E$ symmetry of the eleven-dimensional theory.

\vskip 0.5cm

\subsection{SuperPoincar\'e Algebra in $d=11$}

\vskip 0.3cm

The $SO(2)$ generators stay the same and we introduce the $SO(9)/(SO(2)\times SO(7))$ coset generators
\be
{J^m}\,=\,-\,i\,(\,x\,{\partial^m}\,-\,{x^m}\,{\partial}\,)\,+\-\,{\frac {i}{4\sqrt 2\,\parp}}\;{q}_{\alpha}\,(\,{\gamma^m})^{\alpha\,\beta}\,{q}_\beta
\ee

\be
\bar J^{\;n}\,=\,-\,i\,(\,{\bar x}\,{\partial^n}\,-\,{x^n}\,{\bar \partial}\,)\,+\-\,{\frac {i}{4\sqrt 2\,\parp}}\;{\bar q}_{\alpha}\,(\,{\gamma^n})^{\alpha\,\beta}\,{\bar q}_\beta
\ee
satisying 
\bea
\Big[\,J\,,\,J^m\,\Big]&=&J^m\ ,\qquad \Big[\,J\,,\,\bar J^n\,\Big]~=~-\bar J^n \nonumber \\
\Big[\,J^{pq}\,,\,J^m\,\Big]&=&\delta^{pm}\,J^q\,-\,\delta^{qm}\,J^p \nonumber \\
{\Big[}\,J^m\,,\,\bar J^n\,{\Big ]}&=&\,i\,J^{mn}\,+\,\delta^{mn}\,J,
\eea
where $J\,\equiv\,j$ and the $SO(7)$ generators are
\be
J^{mn}\,=\,-\,i\,(\,x^m\,\partial^n\,-\,x^n\,\partial^n\,)\,-\frac{1}{2 \sqrt 2}\,q^\alpha\, ( \gamma^{mn})^{\alpha \beta}\,�\bar q_\beta.
\ee
\noindent The full $SO(9)$ transverse algebra is generated by $J\,,\,J^{mn}\,,\,{J^m}$ and ${\bar J}^n$. All rotations preserve the chirality of the superfield they act upon. The remaining kinematical generators do not get modified
\bea
J^+\,=\,j^+\ ,\qquad J^{+-}\,=\,j^{+-}\ ,
\eea
while new kinematical generators appear,
\bea
J^{+\,m}&=&i\,x^{m}\,\partial^+_{}\ ; \qquad 
\bar J^{+\,n}~=~i\,\bar x^{n}\,\partial^+_{}\ . 
\eea
The linear part of the dynamical boosts are now 
\bea
J^-_{}&=&i\,x\,\frac{\partial\bar\partial\,+\,{\frac {1}{2}}\,{\partial^m}\,{\partial^m}}{\partial^+_{}} ~-~i\,x^-_{}\,\partial+i\,{\frac { \partial}{\parp}}\,\Big\{\,{ \theta}^\alpha\,{\bar\partial_\alpha}~+~\frac{i}{4\sqrt{2}\,\parp}\,(d^\alpha\,\bar d_\alpha-\bar d_\alpha\,d^\alpha)\,\Big\} \nonumber \\
&&-\,{\frac {1}{4}}\,{\frac {\partial_m}{\parp}}\,{\biggl \{}\,{\parp}\;{\theta^\alpha}\,{{(\,{\gamma^m})}_{\alpha\,\beta}}\,{\theta^\beta}\,-\,{\frac {2}{\parp}}\,\;{\partial^\alpha}\,{{(\,{\gamma^m})}_{\alpha\,\beta}}\,{\partial^\beta}\,+\,{\frac {1}{\parp}}\,\;{d^\alpha}\,{{(\,{\gamma^m})}_{\alpha\,\beta}}\,{d^\beta}\,{\biggr \}}\ . \nn \\
\,
\eea
\noindent The other boosts follow from the $SO(9)/(SO(2)\times SO(7))$ rotations,
\be
J^{-\,m}_{}~=~[\,J^-_{}\,,\,J^{m}_{}\,]\ ;\qquad 
\bar J^{-\,n}~=~[\,\bar J^-_{}\,,\,\bar J^{n}\,]\ .
\ee
The dynamical supersymmetries are thus
\bea
\label{ds}
[\,J^-\,,\,\bar q_{+\,\eta}\,]~\equiv~\nbar{\cal Q}_\eta^{}&=&-\,i\,\frac {\partial}{\parp}}\,{{\nbar q}_{+\,\eta}}\,-\,{\frac{i}{\sqrt 2}}\,{{(\,{\gamma^n}\,)}_{\,\eta\,\rho}}\,q^{~\rho}_{\,+}\,{\frac {\partial^n}{\parp}\ ,\cr
&&\cr
[\,{\bar J}^-\,,\,q_+^\alpha\,]~\equiv~{\cal Q}^\alpha_{}&=&i\,{\frac {\bar \partial}{\parp}}\,{{q_+}^\alpha}\,+\,{\frac {i}{\sqrt 2}}\,{{(\,{\gamma^m}\,)}^{\,\alpha\,\beta}}\,{{\bar q}_{+\,\beta}^{}}\,{\frac {\partial^m}{\parp}}\ ,
\eea
\noindent and satisfy
\bea
\{\,{\cal Q}^{\,\alpha}_{}\,,\,q_+^\eta\,\}~=~-\,{{(\,{\gamma^m}\,)}^{\alpha\,\eta}}\,{\partial^m}\ ,
\eea
\noindent and the supersymmetry algebra,
\bea
\{\,{\cal Q}^{\,\alpha}_{}\,,\,\nbar {\cal Q}^{}_{\,\eta}\,\}~=~i\,\sqrt{2}\,\;{{\delta^{\alpha}}_{\eta}}\,\frac{1}{\parp}
\,\Big(\partial\,{\nbar \partial}\,+\,\frac{1}{2}\,{\partial^m}\,{\partial^m}\,\Big)\ .
\eea
\vskip 0.2cm
\subsection{Dynamical supersymmetry in $d=11$}
\vskip 0.1cm
\ndt If we can construct the dynamical supersymmetry in $d=11$ we have the full theory to this order since we can then construct the Hamiltonian. In principle we should also construct the generators $J^-,\bar J^-,J^{n-}$ and $\bar J^{n-}$. However, it is sufficient to know that the full theory exists and then it is not necessary to explicitly construct these generators (which are always the most difficult ones to construct).  Let us concentrate on $\bar Q_\alpha \phi$ - since this is a gravity theory the dynamical supersymmetry generator must be linear in transverse derivatives to all orders - here we construct it to order $\kappa$. We must consider the following types of terms:

\begin{itemize}
\item Terms with $\bar \partial$. These must be the same as in $d=4$.
\item Terms with $\partial^n$. These are new types of terms associated with the dimensions $4\ldots 10$.
\item Terms with $\partial$. These terms do not exist in the $d=4$ formulation that we have used in the past. The $d=11$ case however, has a $SO(7)$ $R$-symmetry and for a $d=4$ theory with $SO(7)$ $R$-symmetry, instead of $SU(8)$ $R$-symmetry, such a term can appear. 
\end{itemize}
In particular, this means that the oxidized theory cannot be simply reduced to the $d=4$ theory with an $SU(8)$ R-symmetry. This is known from the Cremmer-Julia theory where they had to use duality transformations. Since there are no duality transformations in the light-cone gauge formulation, one instead needs to employ field redefinitions. Notice that when we check chirality and supersymmetry these terms do not talk to each other. This only happens when we invoke the rotations with $J^m$ and $\bar J^n$.
The first type of terms are the known ones from (\ref Q) (ignoring the $\kappa$).
\bea
\bar Q_\alpha{}^{\bar \partial} \phi&= & \frac{1}{\parp}(\bar \partial \bar q_\alpha \phi {\parp}^2 \phi - \parp \bar q_\alpha \phi \parp \bar \partial \phi) \\
&=& \frac{1}{\parp }( E\parp \bar \partial \phi E^{-1} {\parp}^2 \phi)|_{\rho^\alpha}\ ,
\eea 
where
\be E= exp(\frac{\bar q \cdot \rho}{\parp} ).
\ee
By this expression we really mean
\be
\bar Q_\alpha{}^{\bar \partial} \phi = \frac{\delta}{\delta \rho^\alpha} \frac{1}{\parp }( E\parp \bar \partial \phi E^{-1} {\parp}^2 \phi)|_{\rho^\alpha}\ .
\ee
Note that by writing the term using the coherent state-technique its commutation relation with $ q^\alpha$  is automatic. By only using $\bar q$'s chirality is also automatic. Further, the commutation relations with other kinematical generators are obviously correct except the rotations with $J^m$ and $\bar J^n$ - these are the crucial commutators to check.
\vskip 0.1cm
\ndt We can now construct the most general term with $\partial^n$. From the Appendix, we see that terms with three spinors can be written either as a $|8\rangle$ or a $|48\rangle$. For simplicity, we mix them while keeping two free parameters. The most general expression is then 
\bea
\label{new}
\nonumber \bar Q_\alpha{}^{\partial^n} \phi &=&c_1(\gamma^n)^{\beta \gamma} \frac{1}{{\parp}^2}[ E{\parp}^A \partial^n \phi E^{-1}{\parp}^B \phi)|_{\rho^\beta, \rho^\gamma , \rho^\alpha} \\
&+& c_2 (\gamma^n \gamma^m)^{\alpha \delta} (\gamma^{m})^{\beta \gamma}  \frac{1}{{\parp}^2}[ E{\parp}^A \partial^n \phi E^{-1}{\parp}^B \phi)|_{\rho^\beta, \rho^\gamma , \rho^\delta},
\eea
where $A+B=5$ and $c_1$ and $c_2$  are normalization constants to be determined.  Before turning to the third possible term, we calculate the commutator
%\be
%\bar Q_\alpha{}^{\partial} \phi =c_2(\gamma^n)^{\beta \,\gamma} (\gamma^n)^{\delta \,\epsilon} \,\frac{1}{{\parp}^3}}[ E{\parp}^A \partial \phi E^{-1}{\parp}^B \phi]|_{\rho^\beta, \rho^\gamma , \rho^\delta, \rho^\epsilon , \rho^\alpha} ,
%\ee
%where $A+B=7$.
\be
\label{J^n}
[\bar J^m, \bar Q_\alpha] = -\sqrt 2(\gamma^m)_{\alpha  \beta} Q^\beta,
\ee
where
\be
\bar J^{\;n}\,=\,-\,i\,(\,{\bar x}\,{\partial^n}\,-\,{x^n}\,{\bar \partial}\,)\,+\-\,{\frac {i}{4\sqrt 2\,\parp}}\;{\bar q}_{\alpha}\,(\,{\gamma^n})^{\alpha\,\beta}\,{\bar q}_\beta\ .
\ee
By this commutator we really mean
\be
[\delta_{\bar J^n}, \delta_{\bar Q_\alpha}] \,�\phi.
\ee
We compute
\bea \label{c_0}
\nonumber &&[{\frac {i}{4\sqrt 2\,\parp}}\;{\bar q}_{\beta}\,(\,{\gamma^n})^{\beta\,\gamma}\,{\bar q}_\gamma
 ,\, \frac{1}{\parp }( E\parp \bar \partial \phi E^{-1} {\parp}^2 \phi)|_{\rho^\alpha } ] \\ =&&  \frac {i}{4\sqrt 2\,{\parp}^2}\, (\,{\gamma^n})^{\beta\gamma} \,( E{\parp}^2 \bar \partial \phi E^{-1} {\parp}^3 \phi)|_{\rho^\beta\, \rho^\gamma\,\rho^\alpha }\;\;\ .
 \eea
We see from (\ref{new}) that we must choose $A=2$ and $B=3$. Now consider
\bea 
\label{c_1}
\nonumber &&\,[-\,i\,(\,{\bar x}\,{\partial^n}\,-\,{x^n}\,{\bar \partial}\,) \,  , c_1({\gamma^m})^{\beta \,\gamma} \frac{1}{{\parp}^2}[ E{\parp}^2 \partial^m \phi E^{-1}{\parp}^3 \phi)|_{\rho^{\beta}, \rho^{\gamma} , \rho^\alpha}\\
\nonumber &+& c_2 (\gamma^m\gamma^p)^{\alpha \delta} (\gamma^{p})^{\beta \gamma}  \frac{1}{{\parp}^2}[ E{\parp}^2 \partial^m \phi E^{-1}{\parp}^3 \phi)|_{\rho^\beta, \rho^\gamma , \rho^\delta}] \\
\nonumber &=&  \frac{ic_1}{{\parp}^2}({\gamma^n})^{\beta \,\gamma}[ E{\parp}^2 \bar \partial \phi E^{-1}{\parp}^3 \phi)|_{\rho{\beta}, \rho^{\gamma} , \rho^\alpha}\\
&+&  \frac{ic_2}{{\parp}^2}(\gamma^n\gamma^p)^{\alpha \delta} (\gamma^{p})^{\beta \gamma} [ E{\parp}^2 \bar\partial \phi E^{-1}{\parp}^3 \phi)|_{\rho^\beta, \rho^\gamma , \rho^\delta}.
 \eea
 We have to add (\ref{c_0}) and (\ref{c_1}). We see that the terms with $|48\rangle$ must cancel and the term with a $|8\rangle$ is of the correct form in~(\ref{J^n}). We then obtain the first relation 
 \be
 \frac{1}{4\sqrt 2} + c_1 = 0
 \ee
 % Let us now work with the third possible term.
%\begin{align}
%\nonumber \bar Q_\alpha{}^{\partial} \phi &= c_3(\gamma^n)^{\beta \,\gamma} (\gamma^n)^{\delta \,\epsilon} \,\frac{1}{{\parp}^3}[ E{\parp}^3 \partial \phi E^{-1}{\parp}^4 \phi]|_{\rho^\beta, \rho^\gamma , \rho^\delta, \rho^\epsilon , \rho^\alpha} \\
%&+ c_4(\gamma^{mn})^{\beta \,\gamma} (\gamma^{mn})^{\delta \,\epsilon} \,\frac{1}{{\parp}^3}[ E{\parp}^3 \partial \phi E^{-1}{\parp}^4 \phi]|_{\rho^\beta, \rho^\gamma , \rho^\delta, \rho^\epsilon , \rho^\alpha} .
%&\end{align} 
\ndt We continue by performing the commutator
 \begin{align}
\label{46}
\nonumber &[{\frac {i}{4\sqrt 2\,\parp}}\;{\bar q}_{\beta}\,(\,{\gamma^n})^{\beta\,\gamma}\,{\bar q}_\gamma
 ,\,c_1({\gamma^m})^{\delta \,\epsilon} \frac{1}{{\parp}^2}( E{\parp}^2 \partial^m \phi E^{-1}{\parp}^3 \phi)|_{\rho^{\delta}, \rho^{\epsilon} , \rho^\alpha}] \\=&  \frac {ic_1}{4\sqrt 2\,{\parp}^3}\, (\,{\gamma^n})^{\beta\gamma} \,({\gamma^m})^{\delta \,\epsilon} ( E{\parp}^3  \partial ^m \phi E^{-1} {\parp}^4 \phi)|_{\rho^\beta\, \rho^\gamma\,\rho^\delta\, \rho^\epsilon\,\rho^\alpha }\;\;\ .
 \end{align}
 and
\begin{align}
\label{47}
 \nonumber &[{\frac {i}{4\sqrt 2\,\parp}}\;{\bar q}_{\beta}\,(\,{\gamma^n})^{\beta\,\gamma}\,{\bar q}_\gamma
 ,\,c_2 (\gamma^m\gamma^p)^{\alpha \delta} (\gamma^{p})^{\epsilon \eta}  \frac{1}{{\parp}^2}[ E{\parp}^A \partial^m \phi E^{-1}{\parp}^B \phi)|_{\rho^\epsilon, \rho^\eta , \rho^\delta}] \\ =&  \frac {ic_2}{4\sqrt 2\,{\parp}^3}\, (\,{\gamma^n})^{\beta\gamma} \,(\gamma^m\gamma^p)^{\alpha \delta} (\gamma^{p})^{\epsilon \eta}( E{\parp}^3  \partial^m \phi E^{-1} {\parp}^4 \phi)|_{\rho^\beta\, \rho^\gamma\,\rho^\delta\, \rho^\epsilon\,\rho^\eta }\;\;\  .
\end{align}
\ndt We now return to the third possible term
\be 
\bar Q_\alpha{}^{\partial} \phi = c_3(\gamma^n)^{\beta \,\gamma} (\gamma^n)^{\delta \,\epsilon} \,\frac{1}{{\parp}^3}[ E{\parp}^3 \partial \phi E^{-1}{\parp}^4 \phi]|_{\rho^\beta, \rho^\gamma , \rho^\delta, \rho^\epsilon , \rho^\alpha}\;\;\ .
\ee
\ndt Under a rotation, it will contribute with
\begin{align}
\label{49}
\nonumber & \,[-\,i\,(\,{\bar x}\,{\partial^n}\,-\,{x^n}\,{\bar \partial}\,)\, , c_3(\gamma^m)^{\beta \,\gamma} (\gamma^m)^{\delta \,\epsilon} \,\frac{1}{{\parp}^3}[ E{\parp}^3 \partial \phi E^{-1}{\parp}^4 \phi]|_{\rho^\beta, \rho^\gamma , \rho^\delta, \rho^\epsilon , \rho^\alpha} \\
&=  {-ic_3}(\gamma^m)^{\beta \,\gamma} (\gamma^m)^{\delta \,\epsilon} \,\frac{1}{{\parp}^3}[ E{\parp}^3 \partial^n \phi E^{-1}{\parp}^4 \phi]|_{\rho^\beta, \rho^\gamma , \rho^\delta, \rho^\epsilon , \rho^\alpha}\;\;\ .
\end{align}
We now add these three expressions and demand that the result be of the form in~(\ref {Afive}). This calculation is done in the Appendix and the coefficients are found to be
\be
c_1= -\frac{1}{4\sqrt 2}\;\ ,
\ee
\be
c_2= \frac{1}{36\sqrt 2}\;\ ,
\ee
\be
c_3= -\frac{1}{288}\;\ .
\ee
\ndt There is one more term to the rotation
\begin{align}
\nonumber &[{\frac {i}{4\sqrt 2\,\parp}}\;{\bar q}_{\beta}\,(\,{\gamma^n})^{\beta\,\gamma}\,{\bar q}_\gamma
 ,\,c_3(\gamma^m)^{\delta \,\epsilon} (\gamma^m)^{\eta \,\kappa} \,\frac{1}{{\parp}^3}[ E{\parp}^3 \partial \phi E^{-1}{\parp}^4 \phi]|_{\rho^\delta, \rho^\epsilon , \rho^\eta, \rho^\kappa , \rho^\alpha}\\=
 &  \frac {ic_3}{4\sqrt 2\,{\parp}^4}\, ({\gamma^n})^{\beta\,\gamma} 
\gamma^m)^{\delta \,\epsilon} (\gamma^m)^{\eta \,\kappa} \, [ E{\parp}^4 \partial \phi E^{-1}{\parp}^5 \phi]|_{\rho^\beta, \rho^\gamma, \rho^\delta, \rho^\epsilon , \rho^\eta, \rho^\kappa , \rho^\alpha}\;\;\ .
 \end{align}
Using the Appendix (\ref {Aseven}) we can see that the term involving seven $q$'s under Fierz, has the correct form. The correct form for $\bar Q_\alpha$ is then
\begin{align}\label{bar Q}
\nonumber \bar Q_\alpha \phi =&  \frac{1}{\parp }( E\parp \bar \partial \phi E^{-1} {\parp}^2 \phi)|_{\rho^\alpha }\\
\nonumber -& \frac{1}{4\sqrt 2}(\gamma^n)^{\beta \gamma} \frac{1}{{\parp}^2}[ E{\parp}^2 \partial^n \phi E^{-1}{\parp}^3 \phi)|_{\rho^\beta, \rho^\gamma , \rho^\alpha} \\
\nonumber+& \frac{1}{36\sqrt 2} (\gamma^n \gamma^m)^{\alpha \delta} (\gamma^{m})^{\beta \gamma}  \frac{1}{{\parp}^2}[ E{\parp}^2 \partial^n \phi E^{-1}{\parp}^3 \phi)|_{\rho^\beta, \rho^\gamma , \rho^\delta}\\
-& \frac{i}{288 }(\gamma^n)^{\beta \,\gamma} (\gamma^n)^{\delta \,\epsilon} \,\frac{1}{{\parp}^3}[ E{\parp}^3 \partial \phi E^{-1}{\parp}^4 \phi]|_{\rho^\beta, \rho^\gamma , \rho^\delta, \rho^\epsilon , \rho^\alpha}\;\;\ .
\end{align}
\ndt We can now read off  $Q^\alpha$ to be
\begin{align}
\nonumber Q^\alpha \phi = & \frac{i}{72 {\parp}^2}(\gamma^p)^{\alpha \beta} (\gamma^{p})^{\gamma\delta} [ E{\parp}^2 \bar\partial \phi E^{-1}{\parp}^3 \phi)|_{\rho^\beta, \rho^\gamma , \rho^\delta} \\
\nonumber+& \frac{i}{288 \sqrt 2{ \parp}^3}( \gamma^r)^{\alpha \beta} (\gamma^{r})^{\gamma\delta} (\gamma^{m})^{\epsilon\eta} \frac{1}{{\parp}^2}[ E{\parp}^3 \partial^m \phi E^{-1}{\parp}^4 \phi)|_{\rho^\beta, \rho^\gamma , \rho^\delta, \rho^\epsilon,\rho^\eta}\\ 
+& \frac{i}{16128{\parp}^4}(\gamma^r)^{\alpha \beta}(\gamma^r)^{\gamma \delta}(\gamma^m)^{\epsilon \eta} (\gamma^m)^{\kappa \rho} [ E{\parp}^4 \partial \phi E^{-1}{\parp}^5 \phi]|_{\rho^\beta, \rho^\gamma , \rho^\delta, \rho^\epsilon , \rho^\eta, \rho^\kappa, \rho^\rho}\;\;\ .
\end{align}
\ndt We first note that even though these formulae are written using the highly economical light-cone superfield formalism, it would still be very hard to write them had we not used the coherent state technique. We can now ask if this still commutes with the $\E$ transformations in~(\ref{E}). It is easy to see that it will not since the last term in (\ref{bar Q}) contains a bare field $\phi$ which will transform under the constant term in~(\ref{E}) and there is nothing to cancel against it. Hence we conclude that the formulation in this section is not invariant under the transformation (\ref{E}).

\subsection{An oxidized Hamiltonian with $\E$ symmetry}
\vskip 0.1cm
\ndt In an earlier paper~\cite{Ananth:2005vg}, $\mathcal N=8$ supergravity was oxidized to $d=11$, keeping the derivative structure of the $d=4$ theory. The key step  was to introduce the `generalized derivative' 
\bea
{\nbar \nabla}\;=\;{\bar \partial}\,+\,{\frac {\sigma}{16}}\,{{\bar d}_{\alpha}}\,{{(\,{\gamma^m}\,)}^{\alpha\,\beta}}\,{{\bar d}_\beta}\,{\frac {\partial^m}{\parp}}\,,
\eea
which naturally incorporates the coset derivatives $\partial^{m}$ and its partner
\bea
[\;{\nbar \nabla}\,,\,{J^m}\;]\;{\equiv}\;{\nabla^m}\;=\;-\,i\,{\partial^m}\,+\,{\frac {i\,\sigma}{16}}\,{{\bar d}_{\alpha}}\,{{(\,{\gamma^m}\,)}^{\alpha\,\beta}}\,{{\bar d}_\beta}\,{\frac {\partial}{\parp}}\,.
\eea
\ndt Note that the derivative $\partial$ is not introduced. This new derivative $({\nbar \nabla},\nabla^m)$ transforms as a vector under the little group in eleven dimensions. The essential point here, is to keep the eleven-dimensional cubic vertex in the same form as the $d=4$ vertex but with transverse derivatives replaced by generalized derivatives. The cubic vertex is thus 
\bea
\begin{split}
{\mathcal V}\;=&-\;{\frac {3}{2}}\,{\kappa}\;{\int}\,{d^{11}}x\,{\int}\,{d^8}{\theta}\,{d^8}{\bar \theta}\;{\frac {1}{{\parp}^2}}\;{\nbar \phi}\;\;{\nbar \nabla}\,{\phi}\;{\nbar \nabla}\,{\phi}+{\mbox {c.c.}}\ \\ 
\end{split}
\eea
The $SO(2)$ invariance follows from the work in $d=4$ and the $SO(7)$ invariance is covariantly realized so only the invariance under $SO(9)/(SO(7)\times SO(2))$ needs explicit verification. We consider the following variations  
%\bea
%{\delta_{J^m}}\,{\phi}\;=\;\,\frac{i}{\sqrt 2}\;{\omega_m}\,{\parp}\,\;{\theta^\alpha}\,{{(\,{\gamma^m})}_{\alpha\,\beta}}\,{\theta^\beta}\,{\phi}\ ,
%\eea
%\bea
%\begin{split}
%{\delta_{J^m}}\,{\nbar \phi}\,=&\,{\omega_m}\,{\biggl \{}\,{\frac {i}{2\,\sqrt 2}}\,\parp\;{\theta^\alpha}\,{{(\,{\gamma^m})}_{\alpha\,\beta}}\,{\theta^\beta}\,-\,{\frac {i}{\sqrt 2\,%\parp}}\;{\partial^\alpha}\,{{(\,{\gamma^m})}_{\alpha\,\beta}}\,{\partial^\beta}\, \\
%\;\;\;\;\;\;\;\;\;&\;\;\;\;\;\;\;+\,{\frac {i}{2\,\sqrt 2\,{\parp}}}\;{d^\alpha}\,{{(\,{\gamma^m})}_{\alpha\,\beta}}\,{d^\beta}\,{\biggr \}}\,{\nbar \phi}\ ,
%\end{split}
%\eea
\bea
{\delta_{J^m}}\,{\nbar \phi}\,=\frac {i}{2\,\sqrt 2}\, {\omega_m}\, \frac{1}{\parp}\, {q^\alpha}\,{{(\,{\gamma^m})}_{\alpha\,\beta}}\,{q^\beta}\,{\nbar \phi}\, \equiv K({q}) 
\eea
Using the ``inside-out'' constraint (\ref{inside-out}), we obtain
\bea
{\delta_{J^m}}\,{\phi}\;=\;\, \frac{1}{4}\, \frac{(d)^8}{{\parp}^4}\, ({\delta_{J^m}}\, {\nbar \phi})\, \equiv \, \frac{1}{4}\, \frac{(d)^8}{{\parp}^4}\, K({q})\ .
\eea
We also have, from earlier,
\bea
{\delta_{J^m}}\,{\nbar \nabla}\;=\;-\,{\omega_m}\,{\nabla^m}\ ,
\eea
where $\omega_m$ are the parameters of the $SO(9)/(SO(7)\times SO(2))$ coset transformations. This check is straightforward to perform and the only relevant terms in the variation all involve one $SO(2)$ derivative and one $\partial^m$. The net variation yields
\bea
\delta_J\, {\mathcal V}\,\propto\,\int\,{\biggl (}\fr{\sqrt 2}i\sigma+i{\biggr )}\,\fr{{\parp}^2}{\nbar \phi}\,{\bar \partial}\phi\,\partial^m\phi\ ,
\eea
which needs to vanish for invariance under the relevant coset group. This determines $\sigma=-\sqrt 2$ and fixes the generalized derivative entirely. In this light-cone form, the Lorentz invariance in eleven dimensions is automatic once the little group invariance has been established. 
\vskip 0.3cm
\noindent Thus, the ${\mathcal N}=1$ supergravity action, to order $\kappa$, in eleven dimensions is
\be
\label{n=1in11}
\beta\,\int\;d^{11}x\,\int d^8\theta\,d^8 \bar \theta\,{\cal L}\ ,
\ee
where $\beta\,=\,-\,\frac{1}{64}$ and
\bea
\label{eleven}
{\cal L}&=&-\bar\phi\,\frac{\Box}{\partial^{+4}}\,\phi\,-\,2\,\kappa\,(\,\frac{1}{{\parp}^2}\;{\bar \phi}\;\;{\nbar \nabla}\,{\phi}\;{\nbar \nabla}\,{\phi}+\,\frac{1}{{\parp}^2}\;\phi\,\nabla\,{\bar \phi}\,\nabla\,{\bar \phi})\ ,
\eea
\noindent with the d'Alembertian being
\bea
\Box\,=\,2\,(\,\partial\,{\bar \partial}\,+\,\fr{2}\,\partial^m\,\partial^m\,-\,\partial_+\,\parm\,)\ .
\eea
\ndt To investigate a possible $\E$ symmetry of this Hamiltonian we first have to remember that it should be invariant under the maximal subgroup $SU(8)$. We now have to momentarily forget the calculations above and consider the Hamiltonian to be a function of $\theta$ (that transforms as $\bf 8$ of $SU(8)$). The superfields that the Hamiltonian consists of are simply regarded as being built up from representations of $SU(8)$. We can then check that to this order the Hamiltonian is invariant under the $\E$ transformations (\ref{E}). The point being that there are no $\partial$ mixed up with $\bar \partial$ in the expressions in the interaction term to this order.
\vskip 0.1cm
\ndt We know that there is only one supergravity theory in eleven dimensions. Hence the two expressions we have derived must be equivalent. In order to go from one to the other we need to perform a field redefinition. Since we know it must exist we do not need to find it since we will not use it the following discussions. As stated before, we have only performed the calculations to the lowest order in the coupling constant and it would be nice to have this done to higher orders or even to arbitrary order. This cannot be done simply even with sophisticated computer programs. Our experience though tells us that what is true to the lowest order is usually true to all orders.
\vskip 0.1cm
\ndt It should be mentioned that Hohm and Samtleben have, in an interesting series of papers, constructed models with explicit covariant exceptional symmetries. The price they have to pay is to enlarge spacetime. In the case of $E_{7(7)}$ they use a $(4+56)$-dimensional spacetime~\cite{Hohm:2013uia}. Again, since there should be only one supergravity theory in $d=11$ there should be a relation between their model and ours. In this context we also refer the reader to~\cite{new} and references therein. An early approach to studying higher symmetries in $d=11$ involved the introduction of new gauge degrees of freedom~\cite{dN}.

\section{Higher Exceptional Symmetries}
\vskip .1cm

\ndt It was pointed out by Julia after the first paper on $\E$ symmetry that there should exist a maximally supersymmetric supergravity theory in $d=3$ with an $E_{8(8)}$ symmetry. This was subsequently constructed by Marcus and Schwarz~\cite{Marcus:1983hb}, starting from scratch in $d=3$. An alternative way to construct it is to start from the $d=11$ theory and dimensionally reduce to $d=3$~\cite{Nicolai:1986jk}. To find the complete $E_{8(8)}$ theory one has to perform various duality transformations. Since the theory is unique, both approaches should lead to the same theory.
\vskip 0.2cm
\ndt We may use our canonical superfield and construct this theory in the light-cone gauge~\cite{Brink:2008hv}. The $R$-symmetry in this case is $SO(16)$, which is the maximal symmetry that we can represent in the superspace used above.
$$
SO(16)~\supset~ SU(8)\,\times\, U(1)\ ,\qquad {\bf 16} ~=~ \bf8\,+\,\overline{\bf 8}\ .
$$
The $SU(8)$ and $U(1)$ generators are given by

\begin{equation}\label{SU8andU1}
T^i{}_j ~=~ \frac{i}{2\sqrt{2} \,\d^+} \left( q^i \bar q_j\, -\,\frac{1}{8}\,\delta^i{}_j\, q^k \bar q_k \right)\ ,\,\qquad T~=~\frac{i}{4 \sqrt{2} \, \d^+}\,[\, q^{k}_{}\,,\,\bar q_{k}\, ]\ ,
\end{equation}
with commutation relations

\begin{equation*}
 [\,T^i{}_j\,,\,T^k{}_l\,] ~=~\delta^k{}_j\, T^i{}_l - \delta^i{}_l \,T^k{}_j \ ,\qquad
[\,T\,,\,T^i{}_j\,]~=~0\ .
\end{equation*}
The other quadratic combinations describe coset transformations $SO(16)/(SU(8)\times U(1))$

\begin{equation}\label{28}
T^{ij}~=~\frac{1}{2}\frac{1}{\d^+} q^iq^j\, , \qquad T_{ij}~=~\frac{1}{2}\frac{1}{\d^+} \bar q_i \bar q_j\ ,
\end{equation}
which form the ${\bf 28}$ and ${\bf \overline{28}}$ of $SU(8)$, and close on ($SU(8) \times U(1)$) 

\begin{equation*}
\ [\, T^{ij}\,,\, T_{kl}\,  ]~ = ~\delta^j{}_k T^i{}_l \,-\,
\delta^i{}_k T^j{}_l\, -\,\delta^j{}_l T^i{}_k  \,+\,\delta^i{}_l T^j{}_k \, +\, 2\,(\,\delta^j{}_k\delta^i{}_l \,-\,\delta^j{}_l \delta^i{}_k \,)\, T \ .
\end{equation*}
$SO(16)$ acts linearly on the chiral superfield

\[
 \delta^{}_{SU_8} \, \varphi~=~ \omega^j{}_{i}\,T^{i}{}_{j}\, \varphi\ ,~\quad \delta_{U(1)}\, \varphi ~=~ T \,\varphi\ ,
 \]
\begin{equation}
\delta_{\bf 28}\,\varphi~=~ \alpha_{ij}\,\frac{q^iq^j}{\d^+}\, \varphi\ , \qquad 
\delta_{\bf \overline{28}}\,\varphi ~=~ \alpha^{ij} \frac{\bar q_i \bar q_j}{\d^+}\, \varphi \ ,
\end{equation}
where $\omega^j{}_{i}$, $\alpha_{ij}$, and $\alpha^{ij}$  the transformation parameters. Since the superfield is written in terms of $SU(8)\times U(1)$ representations we must also decompose the non-linearly realized quotient group $E_{8(8)}/SO(16)$ into such representations. 
\be
\bf128 = \bf1_{2}'\,+\, \bf28'_{1} \,+\,\bf70_{0}\,+\,\overline{\bf28}'_{-1} \,+\,\bf\bar1'_{-2}\;\;\ .
\ee
We recognize the $\bf 70$ as the representation in  $\E/SU(8)$; the rest of the coset $\E/SO(16)$ transformations form two $U(1)$ singlets, a twenty-eight dimensional representation and its complex conjugate (not to be confused with the $\bf 1$,  $\bf 28$, and $\bf \overline{28}$ in the adjoint representation of $SO(16)$) - all components of the superfield. This means that there is a constant term in the variation of all the components. They all transform as in a $\sigma$-model.
\newline
\ndt  In~\cite{Brink:2008hv} it was found that the
 $E_{8(8)}/SO(16)$ coset transformations could be written in the compact form

\begin{align}\nonumber
&\delta^{}_{\E/SO(16)}\,\phi~=~\frac{1}{\kappa}\,F\,+\,\kappa\,\epsilon^{i_1i_2 \dots i_8}\,\sum_{c=-2}^{2}
\left(\hat{\overline d}_{i_1i_2\cdots i_{2(c+2)}} \partial^{+c}_{}\,F\right)\\
&\quad\times
\Bigg\{ \left(\frac{\partial}{\partial\, \eta} \right)_{i_{2c+5}\cdots i_8}\,\partial^{+(c-2)} \left( e^{\eta\cdot \hat{\bar d} }  \,\partial^{+(3-c)}\phi\, e^{-\eta\cdot \hat{\bar d} } \partial^{+(3-c)}\phi\,\right)\bigg|_{\eta=0}
\,+\, \m O(\kappa^2)\Bigg\},
\end{align}
where the sum is over the $U(1)$ charges $c=2,1,0-1,-2$ of the bosonic fields, and 

\begin{eqnarray}
F&=&\,\frac{1}{{\partial^+}^2}\,\beta\,(y^-)\,\,+\,i\,\theta^{mn}_{}\,\frac{1}{\d^+}\,{\overline \beta}_{mn}\,(y^-)-\,\theta^{mnpq}_{}\,{\overline \beta}^{}_{mnpq}\,(y^-)+\nn \\
&&+\,i\widetilde\theta^{}_{~mn}\,\d^+\,\beta^{mn}\,(y^-)+\,{4}\,\widetilde\theta\,{\d^+}^2\,{\bar \beta}\,(y^-)\ ,\nn
\end{eqnarray}
and
$$\hat{\overline d}_{i_1i_2\cdots i_{2(c+2)}} ~\equiv~ \hat{\overline d}_{i_1}\hat{\overline d}_{i_2}\cdots\hat{\overline d}_{2(c+2)}\ .$$ 
For a more detailed description see~\cite{Brink:2008hv}. It is remarkable that the $E_{8(8)}$ symmetry can also be represented on the same supermultiplet as the $\E$ symmetry. What happens if we dimensionally reduce the $d=4$ theory. Let us start by looking at the equation of motion for the superfield to order $\kappa$. We can obtain this from the action remembering that the superfield is constrained
\be
\label{eom}
\Box\, \phi = \frac{2\kappa}{\partial^+}[ {\bar \partial}^2 \phi \,\,\partial^{2+} \phi - \bar \partial \partial^+ \phi\,\,\bar \partial \partial^+ \phi] + F(\phi\, \bar \phi) + O(\phi^3)\;\;\ .
\ee
The term $F$ is obtained from the complex conjugate of the term in the action leading to the first interaction term. It is clear that a theory invariant under $ E_{8(8)}$ cannot have a three-point coupling since the maximal subalgebra $SO(16)$ (which is the part of the algebra which is linearly realized) will not allow it. The superfield consists of two representations $\bf 128$, one bosonic and one fermonic. That is a spinor representation of $SO(16)$, and we cannot have three spinor representations forming a scalar. This is key to understanding what kind of field redefinition we must make to find the full $ E_{8(8)}$ symmetry. When we dimensionally  reduce to $d=3$ we will have only one transverse derivative which we will write as $\partial$. Let us so dimensionally reduce the equation above and then use the equation of motion to find
\be
\Box \,[\partial^+ \phi \,\, \partial^+�\phi] =   2 [ {\partial}^2 \phi \,\,\partial^{2+} \phi -  \partial \partial^+ \phi\,\, \partial \partial^+ \phi] +  O(\phi^3)\;\;\ ,
\ee
where we have used  the equation of motion  as $ \partial^- \phi = \frac{\partial^2}{\partial^+} \phi +  O(\phi^2)$. We can now rewrite the equation of motion (\ref{eom}) as
\be
\Box\, \phi = \frac{\kappa}{\partial^+}\Box [\partial^+ \phi \,\, \partial^+�\phi]  + F(\phi\, \bar \phi) + O(\phi^3)\;\;\ ,
\ee
and make a field redefinition
\be \phi' = \phi -  \frac{\kappa}{\partial^+} [\partial^+ \phi \,\, \partial^+\phi]\ ,
\ee
to obtain a new equation of motion of the form
\be 
\Box \phi = F(\phi\, \bar \phi) + O(\phi^3)\ .
\ee
\ndt By making systematic field redefinitions like this order by order, one should be able to reach the $ E_{8(8)}$ symmetric formulation. Note that there are no duality transformations  in the light-cone formulation. As mentioned earlier, a duality transformation in the covariant formulation amounts to a field redefinition in the light-cone formulation.

\section{Consequences for discussions about finiteness}
\vskip .1cm

\ndt In principle, we can repeat what we have done for the $d=4$ theory and oxidize the $d=3$ theory. If we do this in a single step up to $d=11$ the full transverse symmetry will need to be  constructed as an $SO(9) \sim SO(9)/SO(8) \times SO(8)$ and the spinors $\theta$ and $\bar \theta$ will transform as the two eight-dimensional spinor representations of $SO(8)$. We can instead, in the spirit of~\cite{Nicolai:1986jk}, keep the $SO(16)$ symmetry and combine the two $\theta$'s into a $\bf 16$ of $SO(16)$. Let us, however, contemplate how this may be done in two steps. First oxidize the $d=3$ theory to $d=4$ and subsequently to $d=11$ as done in previous sections. If we follow the supersymmetry as in section 3.2 we will now have to deal with two kinds of transverse derivatives $\partial$ and $ \bar \partial$. If we follow the first way, concentrating on the dynamical supersymmetry, we will reach the result in (\ref{Q}). We now know that we will loose the explicit $E_{8(8)}$ symmetry. The second approach, oxidation, will focus instead on the Hamiltonian and keep the derivative structure, but adding in $\bar \partial$ appropriately as in section 3.3. We should then be able to maintain the $E_{8(8)}$ symmetry but will obscure the supersymmetry in the process. By a clever field redefinition we should then be able to arrive at this result from the first approach with explicit supersymmetry.
\vskip 0.2cm
\ndt With the construction of a covariant superfield for the four-dimensional theory~\cite{Brink:1979nt}  Howe and Lindstr\"om asked what kind of counterterms could be constructed. They concluded that there should be a possible term at the seven-loop order. In recent years, Bern, Dixon and collaborators~\cite{Bern:2011qn}  have explicitly calculated four-graviton scattering to four loops and found it to be finite. They have even found indications that the loop graphs are ``more" finite than previously expected. We claim that one cannot trust the counterterm arguments in a theory like $(\mathcal N=8,d=4$) supergravity since it has additional symmetries as indicated in this paper. In a covariant formulation, these are often difficult to find since they are non-linearly realized. In the light-cone formulation one can understand the different symmetries but the formalism has to be tailored to make each particular symmetry manifest. As mentioned, there should then exist very complex field redefinitions to go between the various formalisms.
\vskip 0.1cm
\ndt What does all this mean for the finiteness of $(\mathcal N=8,d=4$) supergravity in perturbation theory? It would be rather surprising if the theory were indeed finite but it looks like the only way to really answer this question is through explicit computations. One may try to construct counterterms in the light-cone formulation. This is an arduous task since one has to construct four-point functions with a large number of derivatives. In a previous paper~\cite{Bengtsson:2012dw} it was found that not only do the counterterms have to satisfy the full superPoincar\'e algebra (together with the exceptional symmetry) but they also have to be invariant under residual reparametrization and gauge transformations. For the moment this looks technically very challenging. See also~\cite{new2}.
\vskip 0.1cm
\ndt It has been found that $(\mathcal N=0,d=4$) Yang-Mills theory (pure Yang-Mills theory) and $(\mathcal N=4,d=4$) Yang-Mills as well as $(\mathcal N=0,d=4$) supergravity (Einstein gravity) and $(\mathcal N=8,d=4$) supergravity have ``better" quantum properties than one would expect from counterterm  and power-counting arguments. Interestingly, we have found~\cite{Ananth:2004es} that all these theories have Hamiltonians with very special forms in the light-cone formulation, as quadratic forms,~\cite{Ananth:2006fh}. This quadratic form makes it easier to check various symmetries but we still do not know whether it admits additional symmetries.
\vskip 0.2cm
\ndt It was pointed out by Julia~\cite{Julia:1982gx} that the exceptional symmetry algebra grew bigger under dimensional reduction. In particular, when going from $d=3$ to $d=2$ one should get an $E_9$ infinite symmetry algebra, to $d=1$ an $E_{10}$ symmetry and to $d=0$ an $E_{11}$ symmetry. In more recent time it has been argued by West~\cite{West:2001as} that  the $E_{11}$ symmetry could already be present in the eleven-dimensional theory - see also~\cite{Nicolai:1986jk}. Similarly  Damour, Henneaux and Nicolai~\cite{Damour:2002cu} have argued that the  $E_{10}$ symmetry is present. At this stage we have not been able to to find these symmetries in our formalism but it is clearly very intriguing and well worth further investigation.
\vskip -0.3cm
\begin{center}
* ~ * ~ *
\end{center}

\vskip -0.3cm

\ndt When the action for superstring theory was constructed~\cite{Brink:1976sc}, \cite{Deser:1976rb} the focus was local supersymmetry on the two-dimensional world-sheet. In the corresponding light-cone formulation, one can start with a free two-dimensional action and enlarge it with a full representation of the superPoincar\'e generators in ten dimensions. Can there be additional symmetries lurking behind these structures? This is quite possible. We know that other symmetries, such as U-duality, appear in superstring theory in various dimensions. This is essentially a discretized version of the exceptional symmetries we find in the low-energy limit, the maximal supergravity theory in various dimensions. It is certainly a possibility that there exist even further symmetries, non-linearly realized, present in the theory that make it even better behaved quantum mechanically.
\vskip 0.1cm
\ndt One goal of many of these efforts is to find the magical $M$-theory. Most attempts to find it are based on extensions of the supersymmetric analyses that we are so used to. Are we climbing up the wrong tree? Should one instead be attempting to ascend the `exceptional tree'? Here we can offer no new insights, but our analysis of the symmetries in maximally supersymmetric field theories tell us that we should broaden such investigations.

\vskip -0.7cm
$\,$
\section{Conclusions}
\vskip -0.3cm
We have argued in earlier papers that the $128$ bosons and the $128$ fermions in the superfield (\ref{superfield}) are representations of various superPoincar\'e algebras depending on the dimension of the spacetime we consider. We have seen that they are also representations of the exceptional algebras $\E$ and $E_{8(8)}$. This is best seen in the light-cone gauge formulation where only the physical degrees of freedom are present. This means that both types of symmetries are partially non-linearly realized. In this paper we first reviewed how both these symmetries could be explicitly constructed in the light-cone formulation in $d=4$. However, when we oxidize the theory to eleven dimensions we are forced to choose which of these symmetries to track/follow. Since there is only one theory in eleven dimensions, there must exist a field redefinition to take us from one result to the other. We therefore claim that these exceptional symmetries should make an appearance in all dimensions. 

\ndt We then argued that in four dimensions there are hidden symmetries such as an $E_{8(8)}$ extending the $\E$ that we know of since the first paper on the subject. This suggests that we cannot really trust arguments about perturbative behavior based on the symmetries we know of for the theory. We believe that this is a straightforward explanation of why this theory seems to be more convergent than the arguments based on symmetries and/or power counting indicate. How far this convergence will reach we cannot say, based on our analysis. Only explicit calculations are likely to settle that issue.

\vskip 0.3cm
\ndt {\it {Acknowledgments}}
\vskip 0.1cm

\ndt We thank Marc Henneaux, Axel Kleinschmidt, Mahendra Mali, Hermann Nicolai and Pierre Ramond for helpful discussions. SA and SM acknowledge support from a DST-SERB grant (EMR/2014/000687) and the CSIR NET fellowship (Govt. of India) respectively.

\vskip 1cm
%\begin{center}
%{\bf \LARGE  Appendix} \\
%\end{center}
%\renewcommand{\theequation}{\thesection.\arabic{equation}}

  \renewcommand{\theequation}{A-\arabic{equation}}
  % redefine the command that creates the equation no.
  \setcounter{equation}{0}  % reset counter 
  \section*{APPENDIX}  % use *-form to suppress numbering

\vskip 0.2cm
\subsection*{Spinors and Fierz identities}
\vskip 0.2cm

\ndt We consider $8$-dimensional spinors under $SO(7)$. We have $28$ antisymmetric $\gamma$-matrices. They must be $\gamma^m$ and $\gamma^{mn}$, where $m,n= 1...7$. We will use $\alpha, \beta,.....$ as spinor indices.
\vskip 0.2cm
\ndt
Consider the product of two spinors
\be
\label{spinapp1}
A_\alpha \, A_\beta = -\frac{1}{8} \gamma^m{}_{\alpha�\beta}\, A\gamma^m A + \frac{1}{16} \gamma^{mn}{}_{\alpha�\beta}\, A\gamma^{mn} A. 
\ee
\ndt
Let us now look at an expression
\begin{align}
\nonumber &\gamma^n{}_{�\beta \gamma}\,(E\partial^+{}^2\bar \partial \phi E^{-1}\partial^+{}^3 \phi)|_{\rho^\alpha, \rho^\beta, \rho^\gamma}\\ 
\nonumber
=&\gamma^n{}_{�\beta \gamma} \frac{\partial}{\partial \rho^\alpha} \frac{\partial}{\partial \rho^\beta} \frac{\partial}{\partial \rho^\gamma} \,(E\partial^+{}^2\bar \partial \phi E^{-1}\partial^+{}^3 \phi)|_{\rho=0} \\ \equiv &\gamma^n{}_{�\beta \gamma} \,A_\alpha A_\beta A_\gamma \,\,B.
\end{align}

\ndt
In order to get expressions of the form (40), we now have to Fierz this to obtain a prefactor of the form $\gamma^n{}_{�\alpha \beta}$.
\vskip 0.2cm
\ndt
We will have expressions with up to $7$ A's. So we first have to write all such expressions in terms of their irreducible representations. In the case of two A's we have $28$ different components. They can be written as $\bf{7}+\bf{21}$ as in (\ref {spinapp1}). In the case of three A's we have $56$ components which can be written as $\bf 8$ and $\bf 48$. We  write them as 

\be
\nonumber | 8\rangle_3= (\gamma^q A)_\alpha\,A\gamma^qA,
\ee
\be
\nonumber |48\rangle_3= A_\alpha A\gamma^pA -\frac{1}{7}(\gamma^p\gamma^q A)_\alpha\,A\gamma^qA.
\ee
\vskip 0.2cm
\ndt
We can now decmpose other expressions with $3A$'s in terms of these states such as 
\begin{align}
\nonumber &(\gamma^r A)_\alpha \, A\gamma^{rp}A =\\
\nonumber & 5 ( A_\alpha A\gamma^pA -\frac{1}{7}(\gamma^p\gamma^q A)_\alpha\,A\gamma^qA)\\
-&\frac{2}{7}(\gamma^p\gamma^q A)_\alpha\,A\gamma^qA.
\end{align}
Furthermore by  Fierzing we find that
\be
 ( \gamma^qA)_\alpha \, A\gamma^{q}A  = \frac{1}{2} ( \gamma^{qr}A)_\alpha \, A\gamma^{qr}A,
\ee
\vskip 0.2cm
\ndt
Consider an expression with $4$ $A$'s. There are $70$ independent such terms. What form can they be of? They should be
\vspace{.3cm}

$A\gamma^m A\, A \gamma^{n} A$  $\,\,\,\,$  which is $\bf{1+27}$.
\vspace{.3cm}

$A\gamma^m A\, A \gamma^{mn} A$  $\,\,\,\,$  which is $\bf{7}$.
\vspace{.3cm}

$A\gamma^{[m} A\, A \gamma^{np]} A$  $\,\,\,\,$ 
\vspace{.3cm}
which is $\bf{35}$.

\subsection*{Expressions involving $5$ A's}
\vskip 0.2cm
\ndt
Next we consider expressions with $5A$'s. The corresponding irreducible forms for $5 A$'s are
 \be
\nonumber | 8\rangle_5=  A_\alpha\,A\gamma^rA\, A\gamma^rA
\ee
\be
\nonumber |48\rangle_5=(\gamma^r A)_\alpha A\gamma^pA\, A\gamma^rA -\frac{1}{7}(\gamma^p A)_\alpha\,A\gamma^rA\,A\gamma^rA.
\ee
By Fierzing we can the find the following decomposition.
\begin{align}
\nonumber &A_\alpha \,A\gamma^{pq} A A\gamma^qA =\\
\nonumber -& \frac{2}{3}[(\gamma^r A)_\alpha A\gamma^pA\, A\gamma^rA -\frac{1}{7}(\gamma^p A)_\alpha\,A\gamma^rA\, A\gamma^rA]\\
+&\frac{4}{7}(\gamma^pA)_\alpha\,A\gamma^rA\, A\gamma^rA.
\end{align}
A useful consequence of this formula is 
\be
( \gamma^p A)_\alpha \, A\gamma^{pr}A \,A\gamma^rA = 4 A_\alpha \, A\gamma^{r}A \,A\gamma^rA
\ee
\ndt
Two other useful formulae are 
\begin{align}
\nonumber &A_\alpha \, A\gamma^p A\,A\gamma^qA=\\
\nonumber &\frac{1}{9} (\gamma^p \gamma^r A)_\alpha\, A\gamma^rA\,A\gamma^qA\\
\nonumber +&\frac{1}{9} (\gamma^q \gamma^r A)_\alpha\, A\gamma^rA\,A\gamma^pA\\
+&\frac{1}{9} \delta^{pq}\,A_\alpha\, A\gamma^rA\,A\gamma^pA,
\end{align}
and
\begin{align}
\nonumber &A_\alpha \, A\gamma^{[pq} A\,A\gamma^{r]}A=\\
\nonumber -&\frac{1}{9} (\gamma^{pq} \gamma^s A)_\alpha\, A\gamma^sA\,A\gamma^rA\\
\nonumber -&\frac{1}{9} (\gamma^{rp} \gamma^sA)_\alpha\, A\gamma^sA\,A\gamma^qA\\
-&\frac{1}{9} (\gamma^{qr} \gamma^sA)_\alpha\, A\gamma^sA\,A\gamma^pA.
\end{align}
\ndt
If we add the contributions at the level of $5$ A's we get from (\ref {46}), (\ref {47}) and (\ref {49}), apart from a common prefactor the expression
\begin{align}
\label{Afive}
\nonumber & c_1 A_\alpha \, A\gamma^nA\, A\gamma^mA\, B^m\\
\nonumber +& c_2 (\gamma^m \gamma^pA)_\alpha \, A\gamma^nA\, A\gamma^pA\, B^m\\
\nonumber -& c_3 A_\alpha \, A\gamma^mA\, A\gamma^mA\, B^n\\
\nonumber =&\frac{ c_1}{9} (\gamma^n \gamma^rA)_\alpha \, A\gamma^rA\, A\gamma^mA\, B^m\\
\nonumber +&\left( \frac{ c_1}{9} + c_2 \right) (\gamma^m \gamma^r A)_\alpha \, A\gamma^nA\, A\gamma^rA\, B^m\\
+& \left( \frac{ c_1}{9}- 4 \sqrt{2}\ c_3 \right) A_\alpha\, A\gamma^rA\, A\gamma^rA\, B^n
\end{align}
The first term is of the correct form and the other two have to cancel giving
\be
c_1= -\frac{1}{4\sqrt 2},
\ee
\be
c_2= \frac{1}{36\sqrt 2},
\ee
\be
c_3= -\frac{1}{288}.
\ee
\subsection*{Expressions involving $7$ A's}
\vskip 0.2cm
\ndt
Consider so expressions with seven A's. There is only an $\bf 8$ possible.
\be
|8\rangle_7 = (\gamma^mA)_\alpha\, A\gamma^mA\, A\gamma^nA\, A\gamma^nA.
\ee
\ndt
 By a simple Fierzing we find that
\be
\label{Aseven}
A_\alpha\, A\gamma^n A\, A\gamma^mA\, A\gamma^mA = \frac{1}{7} (\gamma^n \gamma^r A)_\alpha\,A\gamma^rA\,A\gamma^m A\, A\gamma^mA.
\ee

\end{document}